\documentclass[11pt]{article}
\usepackage{moriond,epsfig}

\newcommand{\gev}{\; \hbox{GeV}}

\newcommand{\deu}{(\delta^u_{23})}

\newcommand{\bxs}{ B \rightarrow X_s l^+l^-}
\newcommand{\gcenu}{ B \rightarrow X_c e \nu}
\newcommand{\ds}{\displaystyle}

\renewcommand{\a}{\alpha}

\newcommand{\g}{\gamma}

\newcommand{\G}{\Gamma}

\newcommand{\bea}{\begin{eqnarray}}
\newcommand{\eea}{\end{eqnarray}}
\newcommand{\beq}{\begin{equation}}
\newcommand{\eeq}{\end{equation}}
\newcommand{\nn}{\nonumber}
\newcommand{\fr}{\frac}
\newcommand{\hl}{\hline}

\begin{document}
\vspace*{4cm}
\title{$B \rightarrow X_s \ell^+ \ell^-$ in SUSY}
\author{Enrico Lunghi \footnote{Co-authors: {\bf Antonio Masiero},
                              {\bf Ignazio Scimemi} and {\bf Luca Silvestrini}.}}
\address{\rm         
SISSA-ISAS, Via Beirut 2-4, Trieste, Italy and \\
INFN,  Sezione di Trieste, Trieste, Italy \\ 
E-mail: {\rm lunghi@sissa.it}}
\maketitle
\abstracts{
We study the semileptonic decays $B\rightarrow X_s e^+ e^-$, 
$B\rightarrow X_s \mu^+ \mu^-$ 
in generic supersymmetric extensions of the Standard Model.
SUSY effects are parameterized using the mass insertion approximation formalism
 and differences with MSSM results are pointed out. Constraints on SUSY
contributions coming from other processes ({\it e.g.} $b\rightarrow s \g$) are taken into
account. Chargino and gluino contributions to photon and Z-mediated decays are
computed and non-perturbative  corrections are considered.
We find that the integrated branching ratios  and the asymmetries can
be strongly modified.
Moreover, the behavior of the differential Forward-Backward asymmetry
remarkably changes with respect to the  Standard Model expectation. 
}

\newpage
\section{Introduction}
In this work we want to investigate the  relevance
 of new physics effects in the 
semileptonic inclusive decay $B \rightarrow X_s
l^+ l^-$.
This decay is quite suppressed in the Standard Model;
however, new $B$-factories should  reach the precision requested by the
 SM prediction~\cite{babar} and an estimate of all possible new contributions to
 this process is compelling.

Because of the presence of so many unknown parameters (in particular in the
scalar mass matrices) it is very useful to adopt the so-called 
``Mass Insertion Approximation''(MIA)~\cite{hall}. 
In this framework one chooses a basis for fermion and sfermion states
in which all the couplings of these particles to neutral gauginos are
 flavor diagonal. Flavor changes in the squark sector are provided
 by the non-diagonality of the  sfermion propagators
which are expanded around these off-diagonal entries (Mass Insertions).
A keen analysis of 
the different Feynman diagrams involved will allow us to isolate the few
insertions really relevant for a given process. 
In this way we see that only a small number of the new parameters is
involved and a general SUSY analysis is made possible.

We consider all possible contributions to charmless semileptonic $B$ decays 
 coming from chargino-quark-squark and gluino-quark-squark interactions
and we analyze both Z-boson and photon mediated decays.
Contributions coming from penguin and box diagrams are taken into account;
moreover, corrections to the MIA results due to a light $\tilde t_R$
are considered.
A direct comparison between the SUSY and the SM contributions to
the Wilson coefficients  is performed.
Once   the constraints on mass insertions are established,
we find that in generic SUSY models
 there si still  enough room  in order to see large deviations from
 the SM expectations for branching ratios and asymmetries. 
For our final computation of physical observables
 we consider NLO order QCD evolution of the coefficients and 
non-perturbative corrections ($O(1/{m_b^2}),\ O(1/{m_c^2}),$..),   
each in its proper range of the dilepton invariant mass.

\section{General framework}
\label{sec:opba}
We follow the notation and the conventions of ref.~\cite{bura} for 
what regards the effective Hamiltonian and the operator 
basis. 
 With those definitions the differential branching ratio and 
the forward-backward asymmetry  can be written as
\begin{eqnarray}
R(s) &\equiv& {1\over \G (\gcenu)} {{\rm d} \ \G (\bxs) \over {\rm d} s} \nn \\
&& \hspace*{-1.5cm} = {\a^2 \over 4 \pi^2} \left| K_{ts}
     \over K_{cb} \right| {(1-s)^2
     \over f(z) k(z)} \left[ (1+2s)\left( |\tilde{C}_9^{\rm eff}(s)|^2
      + |\tilde{C}_{10}|^2 \right) +4(1+{2\over s}) |C_7^{eff}|^2 + 
      12 {\rm Re} \left[ C_7^{eff} \tilde{C}_9^{\rm eff}(s) \right] \right],     
\label{eq:br0} \\
A_{FB}(s) & \equiv & {\ds \int_{-1}^1 {\rm d} \cos{\theta} \; {\ds{\rm d}^2 
		     \G (\bxs)\over
                     \ds {\rm d} \cos{\theta} \; {\rm d} s} \; {\rm Sgn} (\cos{\theta})    
                     \over
                     \ds \int_{-1}^1 {\rm d}\cos{\theta} \; {\ds {\rm d}^2 \G (\bxs)\over
                     \ds {\rm d}\cos{\theta} \; {\rm d} s}}  \nn \\
          & = & - \fr{3 {\rm Re} \left[ \tilde{C}_{10} \left(
                    s \ \tilde{C}_9^{\rm eff}(s)  + 2 C_7^{eff} \right)
		     \right]}{ \ds (1+2s)\left( |\tilde{C}_9^{\rm
		     eff}(s)|^2 + |\tilde{C}_{10}|^2 \right)
                 +4(1+{2\over s}) |C_7^{eff}|^2 + 12 {\rm Re} \left[ C_7^{eff}
		     \tilde{C}_9^{\rm eff}(s) \right] } 
\label{eq:afb0} 
\end{eqnarray}
where $s=(p_{l^+}+p_{l^-})^2 /m_b^2$, $\theta$ is
the angle between the positively charged lepton and the B flight
direction in the rest frame of the dilepton system, $f(z)$ and $k(z)$ 
can be found in refs.~\cite{bura1,kim}. 

In the literature the energy asymmetry is also considered~\cite{cho}
but it is easy to show that these two kind of asymmetries are
completely equivalent; in fact a configuration in the dilepton
c.m.s. in which $l^+$ is scattered in the forward direction
kinematically implies $E_{l^+} < E_{l^-}$ in the B rest frame
(see for instance ref.~\cite{alih}).

It is worth underlying that integrating the differential asymmetry
given in eq.~(\ref{eq:afb0}) we do not obtain the global
Foward--Backward asymmetry which is by definition:
\begin{equation}
{N(l^+_\rightarrow) - N(l^+_\leftarrow)
           \over N(l^+_\rightarrow) + N(l^+_\leftarrow)} =
 {\ds \int_{-1}^1 {\rm d} \cos{\theta} \int {\rm d} s\; {\ds{\rm d}^2 
		  \G (\bxs)\over
                     \ds {\rm d} \cos{\theta} \; {\rm d} s} \; {\rm Sgn} (\cos{\theta})    
                     \over
           \ds \int_{-1}^1 {\rm d}\cos{\theta} \int{\rm d} s \; {\ds {\rm d}^2 \G (\bxs)\over
                     \ds {\rm d}\cos{\theta} \; {\rm d} s}} 
\label{intafb2}
\end{equation}
where $l^+_\rightarrow$ and $l^+_\leftarrow$ stand respectevely for
leptons scattered in the forward and  backward direction.

\noindent To this extent it is useful to introduce the following quantity
\beq
\overline A_{FB} (s)  \equiv  {\ds \int_{-1}^1 {\rm d} \cos{\theta} \; {\ds{\rm d}^2 
		     \G (\bxs)\over
                     \ds {\rm d} \cos{\theta} \; {\rm d} s} \; {\rm Sgn} (\cos{\theta})    
                     \over
           \ds \int_{-1}^1 {\rm d}\cos{\theta} \int {\rm d} s \; {\ds {\rm d}^2 \G (\bxs)\over
                     \ds {\rm d}\cos{\theta} \; {\rm d} s}}=
           A_{FB} (s) {R(s) \over \int {\rm d} s R(s)}   
\eeq
whose integrated value is given by eq.~(\ref{intafb2}).

\section{Results}
\label{sec:results}
The genuine SUSY contributions to the coefficients in the most
convenient case ($M_{sq}=M_{gl}=250 \gev $, $M_2=50 \gev$, $\mu=-160
\gev$, $\tan \beta = 2$, $M_{\tilde \nu} =50 \gev$, $M_{\tilde t_R} \simeq 70 \gev$) are
\beq
\cases{C_9^{MI}(M_B) = -1.2 \deu_{LL} + 0.69 \deu_{LR} -0.51(\delta^d_{23})_{LL}      &\cr    
       C_{10}^{MI}(M_B) =  1.75 \deu_{LL} - 8.25 \deu_{LR}     &\cr}
\label{cout}
\end{equation}
where the $\delta$s are the Mass Insertions normalized with
$M_{sq}^2$ ($M_{sq}$ is the avearage squark mass). 

The gluino contributions to $C_7$ are large enough to completely fill
the experimental constraints (coming from the measurement of $BR (b 
\rightarrow s \gamma$):
\beq
0.250 < |C_7^{eff}| <0.445 \ .
\eeq

Taking into account
the  experimental and theoretical limits on the 
$\delta$s we compute the best enhancement and the best depression 
of the SM predictions for the BR and the asymmetries. In addition we 
compute the best enhancement compatible with the conditon of
keeping the SM sign of $C_7^{eff}$.

\begin{figure}
\caption{$R(s)$ [up-left], $A_{FB}(s)$ [up-right], $\overline A_{FB}$
for muons [down-left] and $\overline A_{FB}(s)$ for electrons [down-right]. 
The solid line corresponds to the SM expectation; the dashed and
dotted--dashed lines correspond respectively to the SUSY best
enhancement and depression;
the dotted line is the maximum enhancement obtained without 
changing the sign of $C_7$.}
\begin{center}
\epsfig{file=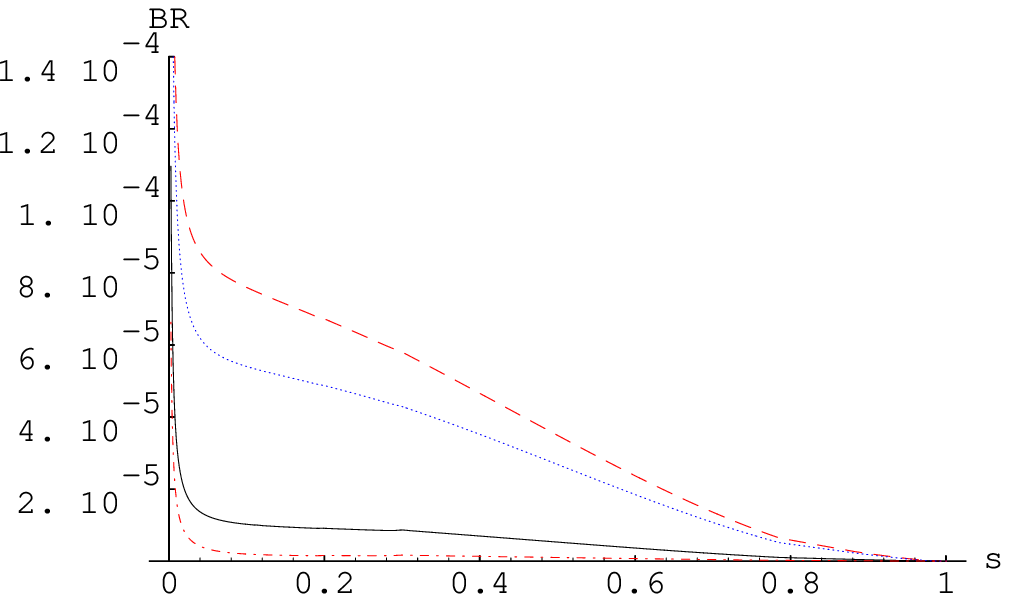,width=0.45\linewidth}
\epsfig{file=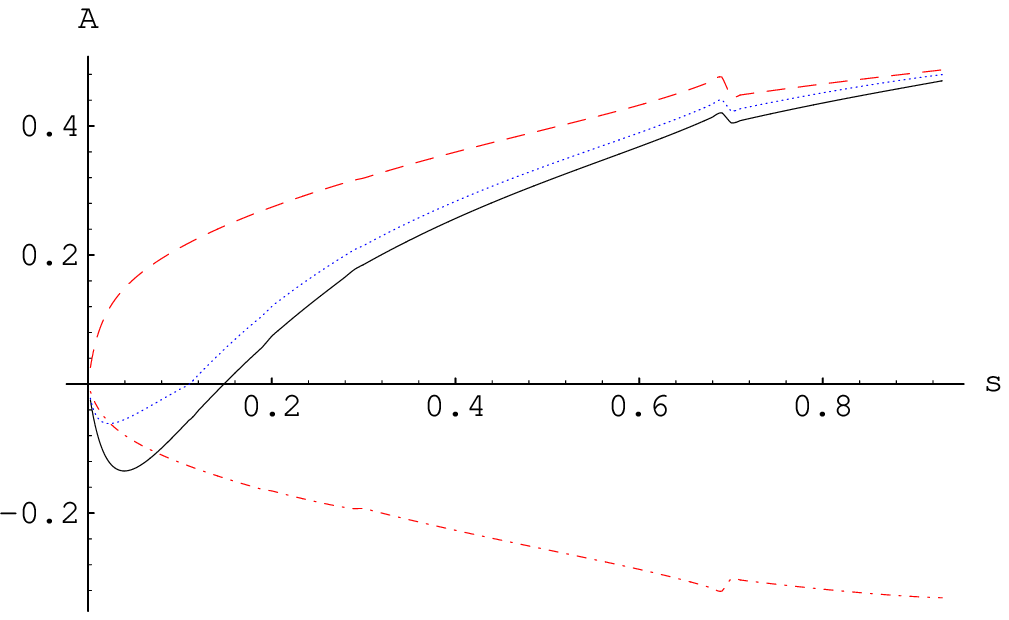,width=0.45\linewidth}
\epsfig{file=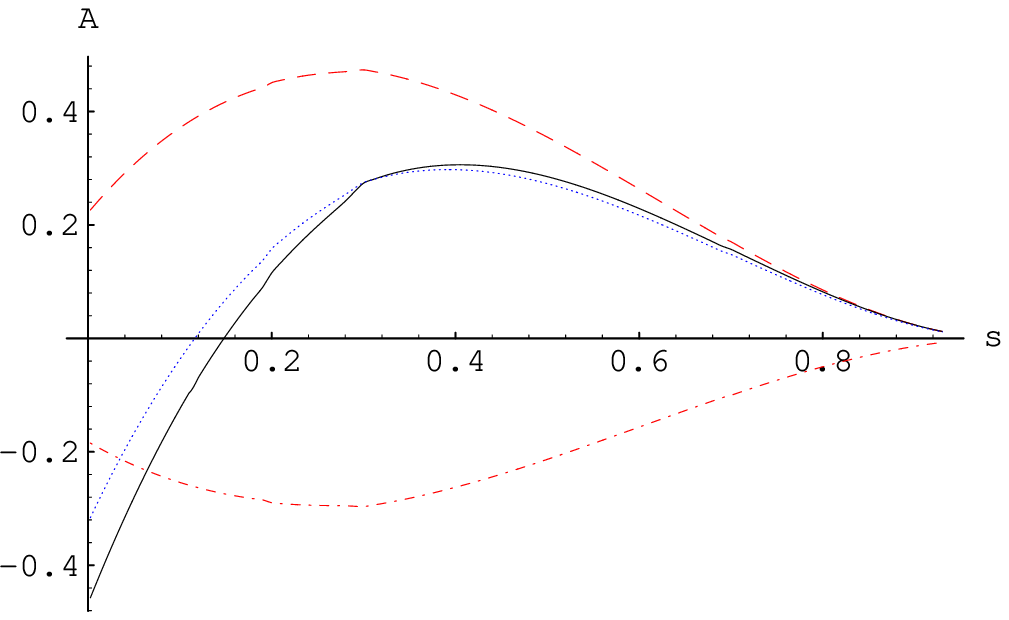,width=0.45\linewidth}
\epsfig{file=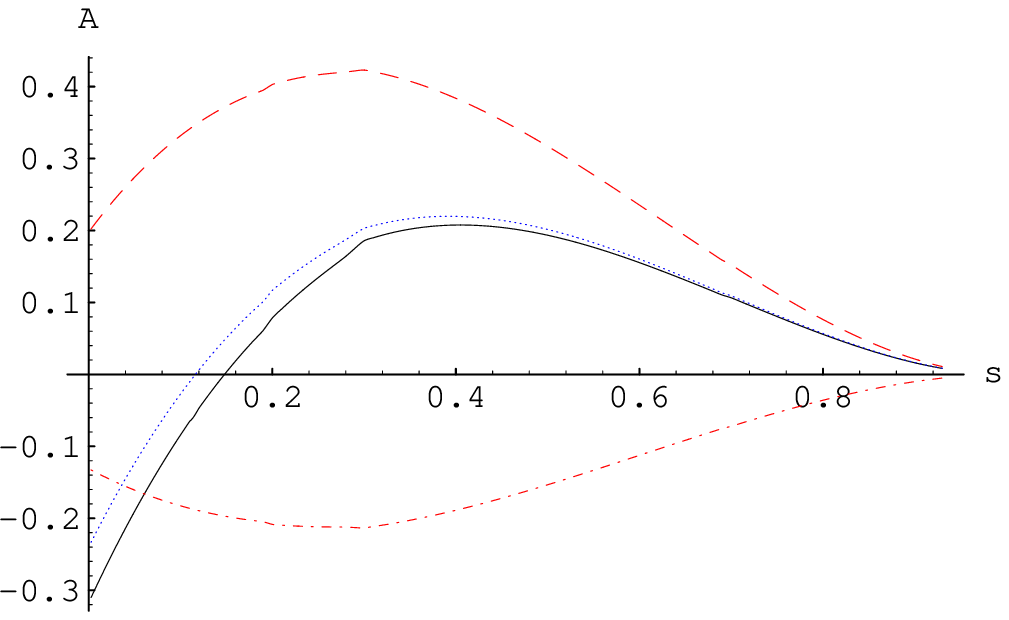,width=0.45\linewidth}
\label{fig:plot}
\end{center}
\end{figure}

The results we obtain are summarized in fig.~\ref{fig:plot} and in 
tab.~\ref{tab:resinteg}. The experimental best limit for the 
BR is~\cite{cleodue} $BR_{exp}< 5.8 \; 10^{-5}$.

\begin{table}
\caption{Integrated BR, $A_{FB}$ and $\overline A_{FB}$ in
the SM and in a general SUSY extension of the SM for the decays
$B \rightarrow X_s e^+ e^-$ and $B \rightarrow X_s \mu^+ \mu^-$. 
The second and third columns
are the extremal values we obtain with a positive $C_7^{eff}$ while
the fourth one is the $C_7^{eff}<0$ case. The actual numerical inputs
for the various coefficients can be found in the text.}
\medskip
\begin{center}
\begin{tabular}{|c|c|c|c|c|c|c|c|} \hline
Observable & SM & SUSY    & SM      &SUSY    & SM      & SUSY      & SM \\  
           &    & maximal & fraction &minimal & fraction & ($C_7<0$) &
           fraction \\ \hline \hline
\vphantom{\fbox{$BR (e)$}}$BR (e)$&$8.6\,{{10}^{-6}}$&$4.3 \
10^{-5}$&$5.0$&$3.2\,{{10}^{-6}}$&$0.37$&$3.4 \ 10^{-5}$&$3.9$  \\ \hl 
\vphantom{\fbox{$BR (e)$}}$A_{FB} (e)$&$0.23$&$0.34$&$1.5$&$-0.22$&$-0.95$&$0.26$&$1.1$  \\ \hl 
\vphantom{\fbox{$BR (e)$}}$\overline A_{FB} (e)$&$0.077$&$0.24$&$3.2$&$-0.12$&$-1.6$&$0.11$&$1.5$  \\ \hline \hline 
\vphantom{\fbox{$BR (e)$}}$BR (\mu)$&$5.8\,{{10}^{-6}}$&$3.8 \
10^{-5}$&$6.5$&$1.4\,{{10}^{-6}}$&$0.24$&$2.8 \ 10^{-5}$&$4.8$  \\ \hl 
\vphantom{\fbox{$BR (e)$}}$A_{FB} (\mu)$&$0.23$&$0.34$&$1.5$&$-0.22$&$-0.95$&$0.26$&$1.1$  \\ \hl 
\vphantom{\fbox{$BR (e)$}}$\overline A_{FB} (\mu)$&$0.11$&$0.27$&$2.4$&$-0.17$&$-1.5$&$0.16$&$1.4$  \\ \hl 

\end{tabular}
\label{tab:resinteg}
\end{center}
\end{table}

\end{document}